\documentstyle[12pt]{article}
\newtheorem{proposition}{Proposition}

\title{Parafermionic and Generalized Parafermionic Algebras}
\author{Dennis Bonatsos${}^a$
\thanks{bonat@cyclades.nrcps.ariadne-t.gr},
\underline{C. Daskaloyannis}${}^b$
\thanks{daskalo@auth.gr} and
 K. Kanakoglou${}^b$ \\[0.15in]
${}^a$
 Institute of Nuclear Physics, NCSR ``Demokritos'',\\
GR-15310 Aghia Paraskevi, Attiki, Greece\\
${}^b$Physics Department, Aristotle University of
Thessaloniki\\
GR-54006 Thessaloniki, Greece}
\date{\phantom{ }}
\begin{document}

\maketitle

\begin{noindent}
{\bf Abstract:}
The general properties of the ordinary and generalized parafermionic algebras 
are discussed. The generalized parafermionic algebras are proved to be 
polynomial algebras. The ordinary parafermionic algebras are shown to be 
connected to the Arik--Coon oscillator algebras.  
\end{noindent}

\bigskip\bigskip

The study of systems of many spins is of interest in many
branches of physics. This study is in many cases facilitated
through boson mapping
procedures (see \cite{Klein} for a comprehensive review). Some
well-known examples are the Holstein--Primakoff mapping of the
spinor algebra onto
the harmonic oscillator algebra \cite{HP} and the Schwinger
mapping of Lie algebras
(or of $q$-deformed algebras) onto the usual (or onto the
$q$-deformed) oscillator algebras \cite{q-oscillator}.

In parallel, in addition to bosons and fermions, parafermions
of order
$p$ have been introduced \cite{Green53,GM64} (with $p$ being a
positive integer), having the characteristic  property that
at most $p$ identical particles of this kind
can be found in the same state. Ordinary fermions
clearly correspond to parafermions with $p=1$, since only one
fermion
can occupy each state according to the Pauli principle. While
fermions
obey Fermi--Dirac statistics and bosons obey Bose--Einstein
statistics
\cite{Isihara}, parafermions are assumed to obey an
intermediate kind
of statistics, called parastatistics
\cite{Isihara,Ohnuki82,Safonov}.The notion
of parafermionic algebras has been recently enlarged by Quesne
\cite{Quesne94}, while the relation between parafermionic
algebras
and other algebras has been given in
\cite{SUSY,Debergh95,Para95}.
The properties of parafermions and parabosons, as well as the
parastatistics and field theories associated with them, have
been the
subject of many recent investigations
\cite{ParaBibl,Macfarlane94}.
Parafermions and parabosons have also been involved in mapping
studies.
A mapping of the spinor algebra onto a parafermionic algebra
has been discussed in \cite{Green53,GM64,Ohnuki82}. Mappings
of so(2n), sp(2n,R), and other Lie algebras onto parafermionic
and parabosonic algebras have been studied in
\cite{Ohnuki82,BisSon88}, while
parabosonic mappings of osp(m,n) superalgebras have been given
in \cite{Palev}.

Recently \cite{ChaiDem96} the algebras of the operators of a 
single spinor
with fixed spin value $j$ have been mapped onto polynomial 
algebras, which
constitute a quite recent subject of investigations in physics
\cite{Quesne94,Para95,LeVi95}. In polynomial algebras the commutator
of two generators does not result in a linear combination of 
generators,
as in the case of the usual Lie algebras, but rather into a 
combination
of polynomials of the generators. The mappings of ref.
\cite{ChaiDem96}
connect the class of spinor algebras to the class of
polynomial algebras.

In the present work we show that the polynomial algebras of
ref. \cite{ChaiDem96}, which are connected to the single spinor
algebras, are indeed examples of either parafermionic algebras
\cite{Green53,GM64,Ohnuki82} or
generalized parafermionic algebras \cite{Quesne94,Para95}.

Let us start by defining the algebra ${\cal A}_n^{[p]}$,
corresponding to
$n$ parafermions of order $p$. This algebra is generated by
$n$ parafermionic generators
$b_i,b_i^\dagger$, where $i=1,2,\ldots ,n$,
 satisfying the {\em trilinear} commutation relations:
\begin{equation}
\left[M_{k\ell}, b_m^\dagger \right] = \delta_{\ell m}
b_{k}^\dagger,
\quad
\left[M_{\ell k}, b_m  \right] = -\delta_{\ell m} b_{k},
\label{eq:ParaDef1}
\end{equation}
where  $M_{k \ell}$ is an operator defined by:
\begin{equation}
M_{k\ell} = \frac{1}{2} \left(
\left[ b_k^\dagger, b_\ell \right] + p \delta_{k\ell} \right).
\label{eq:ParafermionNum}
\end{equation} From
this definition it is clear that eq. (\ref{eq:ParaDef1})
is a {\em trilinear}
relation, i.e. a relation relating three of the operators
$b_i^\dagger, b_i$. Finally the
definition of the parafermionic algebra is completed by the
relation:
\begin{equation}
\left[  b_i, \left[ b_j, b_k \right] \right]=
\left[  b_i^\dagger, \left[ b_j^\dagger, b_k^\dagger \right]
\right]=0.
\label{eq:ParaDef2}
\end{equation}
Each parafermion separately is characterized by the ladder
operators
$b_i^\dagger$ and $b_i$ and the number operator $M_{ii}$.
The basic assumption is that the parafermionic creation and
annihilation operators are nilpotent ones:
\begin{equation}
\left(b\right)^{p+1}=\left(b^{\dagger}\right)^{p+1}=0.
\label{eq:Par0}
\end{equation}
In ref. \cite{SUSY} it is proved that the single parafermionic
algebra is
a generalized oscillator algebra \cite{D1}, satisfying
the following relations (for simplicity we omit the
parafermion indices):
\begin{equation}
\left[ M,b^{\dagger} \right]=b^{\dagger}, \quad
\left[M,b\right]=-b, \label{eq:Par1}
\end{equation}
\begin{equation}
b^{\dagger}b=[M]=M\left(p+1-M\right), \quad
bb^{\dagger}=[M+1]=(M+1)\left(p-M\right), \label{eq:Par2}
\end{equation}
\begin{equation}
M\left(M-1\right)\left(M-2\right)...
\left(M-p\right)=0. \label{eq:Par3}
\end{equation}
The definition (\ref{eq:ParafermionNum}) --~or equivalently
eq. (\ref{eq:Par2})~-- implies the commutation relation:
\begin{equation}
\left[ b^{\dagger},b \right]=2(M-p/2).
\label{eq:CommRel}
\end{equation}
The above relation combined with (\ref{eq:Par1}) suggests the
use of the parafermions as spinors of spin $p/2$
\begin{equation}
S_+ \leftrightarrow b^{\dagger}, \quad
S_- \leftrightarrow b, \quad
S_o  \leftrightarrow (M-p/2).
\label{eq:Spinors}
\end{equation}
The Cayley identity is also valid: 
$$
\prod\limits_{k=- p/2}^{p/2} \left( S_o- k\right)=0.
$$

It is worth noticing that in the case of parafermions the
commutation relation (\ref{eq:CommRel}) is some how trivial
because it is inherent in the definition of the number
operator
(\ref{eq:ParafermionNum}).
This relation switches the trilinear commutation relations to
ordinary commutation relations, where two operators are
involved. In contrast, in the case of parabosons this
construction is
not trivial, because anticommutation relations are involved in
the definition of the number operator\cite{Macfarlane94}.

We start now examining in detail the connection between
spinors with
$j=p/2$ and parafermions of order $p$.

The $p=1$ parafermions coincide with the ordinary fermions,
i.e. the
usual spin 1/2 spinors \cite{BD-Fermi}.

For spinors with $j=1$ Chaichian and Demichev
\cite{ChaiDem96}  use the following mapping
\begin{equation}
S_+ \leftrightarrow \sqrt{2} a^{\dagger}, \qquad\qquad
S_- \leftrightarrow \sqrt{2} a,
\label{eq:Chai-10}
\end{equation}
where
\begin{eqnarray}
a^3={a^{\dagger}}^3=0,   \label{eq:chai-0}  \\
a a^{\dagger} + { a^{\dagger} }^2 a^2= 1. \label{eq:chai-1}
\end{eqnarray}

Using the above two relations we can define the number
operator $N$
\begin{equation}
N=1-\left[ a , a^{\dagger} \right] =
 a^{\dagger} a + { a^{\dagger} }^2 a^2.  \label{eq:chai-2}
\end{equation}
This number operator  satisfies the linear commutation
relations:
$$
\left[N, a^\dagger\right] = a^\dagger,
\quad
 \left[N, a\right] = -a.
$$

The self-contained commutation relations for the $p=2$
parafermions are given
in ref. \cite{Ohnuki82} (eqs (5.13) to (5.20)~)
\begin{eqnarray}
b^3={b^{\dagger}}^3=0,         \label{eq:2para-0}\\
b b^{\dagger} b = 2 b ,     \label{eq:2para-1}\\
b^{\dagger}b^2+ b^2 b^{\dagger}= 2 b. \label{eq:2para-2}
\end{eqnarray}
The set of relations (\ref{eq:2para-0})--(\ref{eq:2para-2})
imply the following definition of the number operator $M$:
\begin{equation}
M=\frac{1}{2} \left( \left[ b^{\dagger}, b \right] +2 \right).
\label{eq:2para-3}
\end{equation}

The set of relations (\ref{eq:chai-0})--(\ref{eq:chai-1}) 
imply
relations (\ref{eq:2para-0})--(\ref{eq:2para-2}) after taking
into consideration the correspondence:
\begin{equation}
a= \frac{1}{\sqrt{2}} b, \qquad
a^{\dagger}= \frac{1}{\sqrt{2}} b^{\dagger}.
\label{eq:2corr}
\end{equation}
For example  one can easily see the following:

\begin{itemize}
\item[a)] Eq. (\ref{eq:2para-0}) occurs trivially from eq.
(\ref{eq:chai-0}).
\item[b)] Eq. (\ref{eq:2para-1}) is obtained by multiplying
eq. (\ref{eq:chai-1}) by $a$ on the right and using eq.
(\ref{eq:chai-0}).
\item[c)] Eq. (\ref{eq:2para-2}) is obtained by multiplying
eq. (\ref{eq:chai-1}) by $a$ on the left and using eq.
(\ref{eq:chai-0}) and (\ref{eq:chai-1}).
\end{itemize}

In ref. \cite{SUSY} the parafermionic algebra
(\ref{eq:2para-0})--(\ref{eq:2para-3}) was shown to be
equivalent to the deformed oscillator algebra \cite{D1}, which
is defined by relations (\ref{eq:Par0})--(\ref{eq:Par3}), for
$p=2$. This deformed oscillator algebra satisfies in addition 
the
relations (\ref{eq:chai-0}) to (\ref{eq:chai-2}). Therefore
the Chaichian - Demichev polynomial algebra
(\ref{eq:chai-0})--(\ref{eq:chai-2}), the $p=2$ parafermionic
algebra (\ref{eq:2para-0})--(\ref{eq:2para-3}) and the 
deformed
oscillator algebra (\ref{eq:Par0})--(\ref{eq:Par3}) are
equivalent.

Relations (\ref{eq:chai-1}) and (\ref{eq:chai-2}) indicate
that $a a^{\dagger}$ and $N$ can be expressed as a linear
combination of monomials $\left(a^{\dagger}\right)^k a^k$.
This is the
reason the algebra described by eqs
(\ref{eq:chai-1})--(\ref{eq:chai-2})
is called in \cite{ChaiDem96} a ``polynomial'' algebra.

What we have just seen is that the polynomial
algebra  (\ref{eq:chai-0})--(\ref{eq:chai-2}) is in fact the
$p=2$ parafermionic
algebra (\ref{eq:2para-0})--(\ref{eq:2para-3}). The new result
which arises from
this discussion is that the parafermionic algebra can be
written as a
polynomial algebra through the r.h.s of eq. (\ref{eq:chai-2}).
It seems that this fact has been ignored, while the ``dual''
relation, giving $b^{\dagger}b$ or $b b^{\dagger}$ as
polynomial functions of the number operator,
$$
b^{\dagger}b= M(3-M), \qquad
b b^{\dagger} = (M+1) (2-M),
$$
is known \cite{Quesne94,SUSY,Debergh95}.

For spinors with $j=3/2$ Chaichian and Demichev
\cite{ChaiDem96}
use  the following mapping
\begin{equation}
S_+ \leftrightarrow \sqrt{3} a^{+}, \quad
S_- \leftrightarrow \sqrt{3} a,
\label{eq:Chai-26}
\end{equation}
where
\begin{eqnarray}
a^4={a^{\dagger}}^4=0, \\  \label{eq:3chai-0}
a a^{\dagger} =1+\frac{1}{3}a^{\dagger}  a
-\frac{1}{3} { a^{\dagger} }^2 a^2 -
\frac{2}{3} { a^{\dagger} }^3 a^3,\\ \label{eq:3chai-1}
\left[ a , a^{\dagger} \right] = 1-\frac{2}{3}N. \\
\label{eq:3chai-2}\end{eqnarray}
The last two equations imply the following expansion of the number 
operator:
\begin{equation}
N = a^{\dagger} a + \frac{1}{2}{ a^{\dagger} }^2 a^2
+  { a^{\dagger} }^3 a^3. \label{eq:3chai-3}
\end{equation}
These relations are the analogues of eqs.
(\ref{eq:chai-0})--(\ref{eq:chai-2}) for the
$j=3/2$ case.

The complicated self-consistent commutation relations for the
$p=3$ parafermionic algebra are given in Appendix B of ref.
\cite{Ohnuki82}. After long but straightforward
calculations the $p=3$ parafermionic relations are deduced
from the above eqs
(\ref{eq:3chai-0})--(\ref{eq:3chai-3}) by taking into account
the correspondence:
\begin{equation}
a= \frac{1}{\sqrt{3}} b, \quad
a^{\dagger}= \frac{1}{\sqrt{3}} b^{\dagger}.
\label{eq:3corr}
\end{equation}
 Therefore the polynomial algebra
(\ref{eq:3chai-0})--(\ref{eq:3chai-3}) is in
fact the $p=3$ parafermionic algebra. The new result which
again
arises from this discussion is that the parafermionic algebra
can be
written as a polynomial algebra through eq.
(\ref{eq:3chai-3}), while
 the ``dual'' relation
$$
b^{\dagger}b= M(4-M), \quad
b b^{\dagger} = (M+1) (3-M),
$$
is again already known \cite{SUSY}.

Stimulated by the above results we can show the following
proposition:

\begin{proposition}
The  {$j=p/2$} spinor algebra $\left\{ S_\pm, S_o \right\}$ is
mapped onto the
  $p$-parafermionic algebra $\left\{b^{+},b, M\right\}$
 which is a polynomial algebra given by the relations:
\begin{equation}
\begin{array}{c}
\left[M, b^{\dagger} \right] = b^{\dagger},\\
\left[M, b \right] = -b, \\
b^{p+1}=\left(b^{\dagger}\right)^{p+1}=0, \\
b^{\dagger}b= M\left(p+1-M\right)= \left[ M \right],\\
b b^{\dagger}=
\left(M+1\right)\left(p-M\right)=\left[ M +1\right], \\
M=\frac{1}{2}\left( \left[b^{\dagger},b\right] +p \right),\\
\end{array}
\label{eq:PolyPara}
\end{equation}
where the number operator $M$ is given by the following
polynomial relation
\begin{equation}
M=\sum\limits_{k=1}^{p}\,
\frac{c_k}{p^k}\, {b^{\dagger}}^k b^k.
\label{eq:new}
\end{equation}
With the ``factorial'' $\left[k\right]!$ being defined as
$$
\left[0\right]!=1, \quad 
\left[n\right]!=\left[n\right] \left[n-1\right]!=
\prod\limits_{\ell=1}^{n} 
{\left[\ell\right]}
=\frac{n! p!}{(p-n)!}, 
$$
the coefficients $c_1, \, c_2 ,\, \ldots , c_p$ can be
determined from
the solution of the system of equations:
\begin{equation}
\left.
\begin{array}{l}
\rho(1)=1\\
\rho(2)=2\\
\ldots \\
\rho(p)=p
\end{array}
\right\}
\label{eq:Nsytem}
\end{equation}
where
$$
\rho(n)= \sum\limits_{k=1}^{n} \frac{c_k}{p^k}
\frac{\Gamma(n+1)}{\Gamma(n-k+1)}  \frac{\Gamma(p+k-n+1)}{\Gamma(p-n+1)}.
$$
\end{proposition}
This is true because we can see that
$$
{b^{\dagger}}^k b^k = \prod\limits_{\ell=0}^{k-1}\left[ M-\ell\right]
\equiv
\frac{\Gamma(M+1)}{\Gamma(M-k+1)}  \frac{\Gamma(p+k-M+1)}{\Gamma(p-M+1)}.
$$
The fact that the number operator of a
parafermionic algebra can be written as a combination of
monomials, i.e.  eq. (\ref{eq:new}), was not
previously known in the context of parafermionic algebras. 
The polynomial expressions as in eq. (\ref{eq:new}) are similar to the
ones used for the construction of the projection operators in the case of
the su(2) algebra \cite{Loedwin,Asherova}. The projection operator method 
has also been used in the case of the su$_q$(2) and  su$_q$(1,1) algebras
\cite{Smirnov90} as a dynamic tool for the calculation of the Clebsch-Gordan 
coefficients. On the other hand the parafermionic algebra is a 
finite dimensional realization of the su(2) algebra, coinciding with the 
spinor algebra.
 
 The analytic calculation
of the coefficients $c_k$ can be achieved by expanding the number operator 
$M$ in a sum over the projection operators
$$
P_m \left\vert n \right> = \delta_{nm} \left\vert n \right>
$$
in the following way:
$$
M= \sum\limits_{m=1}^{p} m P_m
$$
The projection operator $P_o$ to the lowest weight eigenvalue
is given by the expression:
$$
P_o = 
\sum\limits_{k=0}^{p}
{ d_k \left(b^+\right)^k b^k }, 
$$
while 
$$
P_m = \frac{1}{ \left[m\right]!}\left(b^+\right)^m  P_o b^m
=\frac{1}{ \left[m\right]!}
\sum\limits_{k=0}^{p-m}
{ {d_k} \left(b^+\right)^k b^k }, 
$$
where the coefficients $d_n$ are given by the recurrence formula
$$
\begin{array}{l}
d_0=1, \\
d_n= - \sum\limits_{k=0}^{n-1} 
\frac{d_k}{ \left[ n-k \right]! }. 
\end{array}
$$
Then the general solution is given by:
\begin{equation}
d_n =
\sum\limits_{i=1}^{n} 
{ \left( -1 \right)^i\left(
\sum\limits_{ 
\begin{array}{c}
0 <k_1,k_2, \ldots ,k_i \leq n\\
k_1+k_2+\ldots+k_i=n
\end{array} }
\frac {1}{ \left[k_1\right]!
\left[k_2\right]!\ldots \left[k_i\right]!}
\right)}
\label{eq:proj-coef}
\end{equation}

We must point out that these formulae are not specific to the chosen 
parafermion structure function
$\left[ x \right] = x (p+1- x)$ 
and  can be applied for any parafermionic oscillator stucture function.

The number operator $M$ can be expressed using the projection operators:
$$
M = \sum\limits_{m=1}^p m P_m =
\sum\limits_{k=1}^p \frac{c_k}{p^k}  \left(b^+\right)^k b^k , 
$$
while the coefficients $c_n$ can be fould to be  
$$
c_n= p^n \sum\limits_{k=1}^n \frac{k}{ \left[ k \right]! } d_{n-k}. 
$$
In
 Table 1
the coeffients up to $p=5$ are explicitly given.

\begin{table}[ht]
\begin{center}
Table 1:  Coefficients appearing in  eq. (\ref{eq:new}).
$$
\begin{array}{c | c c c c c }
\hline
p &  c_1  &  c_2  & c_3& c_4  & c_5 \\
\hline\hline
\displaystyle
 1 &  1  & --  & -- & -- & -- \\
\displaystyle
 2 &  1     & 1 & -- & -- & -- \\
\displaystyle
 3 &  1     & 1/2  & 1 & -- & -- \\
\displaystyle
 4 &  1     & 1/3
  &  1/3 & 7/9 & -- \\
\displaystyle
 5 &  1     & 1/4 & 1/6
  &19/96 & 23/48\\
\hline
\end{array}
$$
\end{center}
\end{table}

One must notice that the parafermionic algebra 
(\ref{eq:Par1}--\ref{eq:Par3}) has affinities with the Arik -- Coon 
$Q$-deformed algebra \cite{Arik}, which is defined by the relations:
\begin{equation}
[N, a]= -a, \quad [N, a^{\dagger}]= a^{\dagger}, \quad
a^{\dagger} a = [N]_Q, \quad  a a^{\dagger}  = [N+1]_Q
\label{eq:Arik}
\end{equation}
where $[x]_Q = (1- Q^x)/(1-Q)$. The generators of this oscillator satisfy the
following commutation relation
\begin{equation}
[a,a^{\dagger}]= Q^N
\label{eq:ArikCR}
\end{equation}
By defining $Q = \exp[- \tau]$ the commutation relation (\ref{eq:ArikCR}) can
be written, for $\tau \to 0$ as
\begin{equation}
[a,a^{\dagger}]= \exp[-\tau N] = 1 - \tau N + {\cal O}(\tau^2)
\label{eq:eq:ArikCRApprox}
\end{equation}
After comparing the above equation with equation (\ref{eq:CommRel}), we can see
that there is a approximative mapping of the parafermionic oscillator to the 
Arik -- Coon oscillator by putting
$$
\begin{array}{ccc}
\hline
\rm{Arik-Coon} & \to & \mbox{para Fermi}
\\  \hline
b  & \to &\sqrt{p} \; a \\
b^{\dagger} & \to &  \sqrt{p} \; a^{\dagger} \\
M & \to & N \\
\tau &= & 2/p
\end{array}
$$ 
The meaning of the order $p$ of the parafermionic oscillator is quite clear, 
$p$ is a the ''{\em capacity}'' of the oscillator, i.e. the maximum number of 
permitted states, which can be found similtaneously at the same position. 
Therefore, the parameter $Q$ of the Arik -- Coon oscillator is a ''{\em 
measure }'' of the similtaneously existed states at the same position.

 A nice example is the case of the $J=0$ pairing of nucleons in a closed 
nuclear shell. The algebra of the fermion pairs coupled to angular momentum 
zero are descibed by the algebra \cite{Klein}:
$$
\left[ A_0, A_0^{+} \right] = 1 -N_F/\Omega,
\quad
\left[ \frac{N_F}{2}, A_0^{+} \right] =A_0^{+},
\quad
\left[ \frac{N_F}{2}, A_0 \right] = - A_0
$$
where $N_F$ is the number of fermions, $2 \Omega = 2 j +1 $ is the size of 
the shell, i.e. the "capacity" of our space.
The simplest piring Hamiltonian is given by:
$$
H = - G \Omega A_0^{+} A_0
$$

 For the above algebra there is a natural mapping to parafermions of order $p$,
each parafermion corresponds to a Fermi pair \cite{BD-nuclear} and 
$p = \Omega$. The ordinary $q$-deformed oscillator \cite{q-oscillator} fails 
to give an approximation of the pairing model, while the Arik -- Coon 
oscillator is quite satisfactory \cite{Bonat}.

\bigskip\bigskip

{\it Conclusions} The
parafermionic algebras can be considered as polynomial
algebras, their diagonal number operator $M_{ii}$ being able
to be written as a combination of monomials of the ladder
operators. The general problem of finding an expression of the
number operator $M_{ij}$ as a combination of monomials of the
ladder operators is still open. A similar problem
exists in quonic algebras \cite{quons,ChaiFM93,LuZhe94}.
Work in this direction is in progress. The Arik -- Coon deformed oscillator 
is a fair approximation of the parafermionic algebra.

Support from the Greek Secretariat of Research and Technology
under contract PENED 95/1981  is gratefully acknowledged.

\end{document}